\documentclass[a4paper,fleqn,usenatbib]{mnras}

\usepackage{mathptmx}

\usepackage[T1]{fontenc}
\usepackage{ae,aecompl}

\usepackage{graphicx}	
\usepackage{amsmath}	
\usepackage{amssymb}	

\newcommand{\cmnt}[1]{}

\newcommand{\review}[1]{#1}

\newcommand{\warwick}{$^{1}$}

\newcommand{\ioacambs}{$^{2}$}
\newcommand{\iac}{$^{5}$}

\newcommand{\dpmacal}{$^{7}$}
\newcommand{\lalaguna}{$^{6}$}
\newcommand{\kavli}{$^{3}$}
\newcommand{\sbarb}{$^{4}$}

\newcommand{\hei}{\ion{He}{I}}
\newcommand{\heii}{\ion{He}{II}}
\newcommand{\siii}{\ion{Si}{II}}

\newcommand{\kwd}{$13.8 \pm 3.2$\,km/s}
\newcommand{\phiwd}{$4 \pm 15$\,$^\circ$}

\title[Phase-Resolved Spectroscopy of Gaia14aae]{Phase-Resolved Spectroscopy of Gaia14aae: Line Emission From Near the White Dwarf Surface}

\author[M. J. Green et al.]{
M. J. Green\warwick\thanks{E-mail: matthew.green@warwick.ac.uk (MJG)} 
T. R. Marsh\warwick,
D. Steeghs\warwick,
E. Breedt\ioacambs,
T. Kupfer\kavli$^,$\sbarb,
\newauthor
P. Rodr\'{i}guez-Gil\iac$^,$\lalaguna,
J. van Roestel\dpmacal, 
R. P. Ashley\warwick,
L. Wang\warwick,
E. Cukanovaite\warwick,
\newauthor
and S. Outmani\warwick.
\\
\warwick Astronomy and Astrophysics Group, Department of Physics, University of Warwick, Coventry, CV4 7AL, United Kingdom
\\
\ioacambs Institute of Astronomy, University of Cambridge, Madingley Road, Cambridge, CB3~0HA, United Kingdom
\\
\kavli Kavli Institute for Theoretical Physics, University of California, Santa Barbara, CA 93106, USA 
\\
\sbarb Department of Physics, University of California, Santa Barbara, CA 93106, USA
\\
\iac Instituto de Astrof\'{i}sica de Canarias, E-38205 La Laguna, Tenerife, Spain
\\
\lalaguna Departamento de Astrof\'{i}sica, Universidad de La Laguna, E-38206 La Laguna, Tenerife, Spain
\\
\dpmacal Division of Physics, Mathematics and Astronomy, California Institute of Technology, Pasadena, CA 91125, USA
}

\date{Accepted XXX. Received YYY; in original form ZZZ}

\pubyear{2015}

\begin{document}
\label{firstpage}
\pagerange{\pageref{firstpage}--\pageref{lastpage}}
\maketitle

\begin{abstract}
AM\,CVn binaries are a class of ultracompact, hydrogen-deficient binaries, each consisting of a white dwarf accreting helium-dominated material from a degenerate or semi-degenerate donor star.
Of the 56 known systems, only Gaia14aae undergoes complete eclipses of its central white dwarf, allowing the parameters of its stellar components to be tightly constrained.
Here, we present phase-resolved optical spectroscopy of Gaia14aae.
We use the spectra to test the assumption that the narrow emission feature known as the `central spike' traces the motion of the central white dwarf.
We measure a central spike velocity amplitude of $13.8 \pm 3.2$\,km/s, which agrees at the 1\,$\sigma$ level with the predicted value of $17.6 \pm 1.0$\,km/s based on eclipse-derived system parameters. The orbital phase offset of the central spike from its expected position is $4 \pm 15$\,$^\circ$, consistent with 0\,$^\circ$. 
Doppler maps of the \hei\ lines in Gaia14aae show two accretion disc bright spots, as seen in many AM\,CVn systems. 
The formation mechanism for the second spot remains unclear.
We detect no hydrogen in the system, but we estimate a 3 $\sigma$ limit on H$\alpha$ emission with an equivalent width of -1.14\,\AA.
Our detection of nitrogen and oxygen with no corresponding detection of carbon, in conjunction with evidence from recent studies, mildly favours a formation channel in which Gaia14aae is descended from a cataclysmic variable with a significantly evolved donor.

\end{abstract}




\section{Introduction}
\label{sec:introduction}

AM\,CVn-type binaries are accreting, ultracompact systems with orbital periods of 5--65\,minutes \citetext{see \citealp{SolheimAMCVn}, \citealp{Breedt2015} and \citealp{Ramsay2018} for reviews}.
Each system consists of a white dwarf accreting hydrogen-deficient, helium-dominated matter from an evolved donor, which may be degenerate or semi-degenerate. 
In most AM\,CVn systems, this mass transfer occurs via an accretion disc around the central white dwarf.
AM\,CVn binaries are strong, galactic sources of gravitational waves \citep{Korol2017,Kupfer2018,Breivik2018}. 
They share many properties with hydrogen-accreting cataclysmic variables (CVs), including their accretion geometries and the occurrence of dwarf nova outbursts (dramatic changes in magnitude due to thermal instabilities in the accretion disc). 



The number of AM\,CVn binaries known has increased significantly in recent years, and now stands at 56 \citep{Ramsay2018}. 
The bottleneck in our understanding of AM\,CVns is no longer the small number of known systems, but the difficulty in constraining the properties of many of them.
Key properties of interest in the characterisation of AM\,CVns are their orbital periods and their component stellar masses, as these properties constrain their evolutionary history and formation mechanism. 
However, because of the faintness of the donor star relative to the rest of the system no AM\,CVn donor has been directly observed, rendering mass measurements difficult.


There are three methods by which the stellar mass ratios ($q = M_2 / M_1$, donor mass over accretor mass) of AM\,CVns have been determined: by eclipse photometry, by spectroscopy, or using the `superhump' method. 
The eclipse method uses photometry of the eclipse of the central white dwarf to determine the dynamics and masses of the two stars. It is the most precise method of determining stellar masses, but it is limited to the two known AM\,CVns in which the central white dwarf is fully or partially eclipsed \citep{Copperwheat2011,Green2018}.
The spectroscopic method relies on resolving the radial velocity of the central white dwarf as a function of orbital phase, thereby determining the dynamics of the binary. It therefore requires a spectroscopic signature of the central white dwarf \citep{Roelofs2005a,Kupfer2016}.
Lastly, the superhump method is based on an empirical relation between mass ratio and a photometric signal known as the `superhump period' which is related to disc precession \citep{Patterson1993,Patterson2005,Knigge2006,Kato2013} \cmnt{+ McAllister2018 if published?}. The superhump relation was derived for hydrogen-dominated cataclysmic variables, and although the relation is widely used for AM\,CVns, there are some doubts about its applicability due to the hydrogen-deficient nature of the discs and the extreme values of $q$ found in AM\,CVn binaries \citep{Pearson2007,Kato2014}.

Due to the limited number\footnote{\review{To date, two AM\,CVn binaries have measurements of $q$ by eclipse photometry, four by spectroscopy, and nine by the superhump method \citep[][and references therein]{Green2018b}.}}
of systems for which each method can be applied, there are very few AM\,CVn binaries for which results from multiple methods can be directly compared. The first known AM\,CVn-type binary, AM\,CVn itself, has been studied using both spectroscopic and superhump methods, producing discrepant measurements of $q = 0.18 \pm 0.01$ and $0.101 \pm 0.005$ respectively \citep{Skillman1999,Roelofs2006}. On the other hand, the partially-eclipsing binary YZ\,LMi was studied with both eclipse and superhump methods, producing consistent values of $q = 0.041 \pm 0.002$ and $0.049 \pm 0.008$ \citep{Copperwheat2011}\footnote{Several versions of the empirical superhump relationship are available. For both superhump-derived mass ratios quoted in this paragraph we have used the relationship of \citet{Knigge2006}.}.
Aside from these two systems, there have been no opportunities to test the three methods against each other for AM\,CVns.

Spectral lines of AM\,CVn-type binaries have orbital phase-dependent profiles arising from the presence of multiple components, which we describe in the following two paragraphs.
The accretion disc contributes a broad, double-peaked line profile, as the approaching limb of the disc is blueshifted and the receding limb is redshifted.
The point of contact between the stream of infalling matter and the accretion disc, a region known as the `bright spot', is the origin of a spectral component whose Doppler shift varies with orbital phase (known as an `S-wave').
Many AM\,CVns show a `second bright spot' spectral feature which is Doppler shifted with a similar amplitude, but offset in orbital phase by 120$^\circ$ from the first bright spot \citep{Roelofs2005a,Roelofs2006,Kupfer2013,Kupfer2016}. The physical cause of this feature is uncertain. 
Similar features have been seen in some short-period hydrogen CVs, though it is unclear whether the mechanism is the same \citep[eg.\ EX\,Dra,][]{Joergens2000}.

Lastly, some AM\,CVn spectral lines show a sharp emission feature known as the `central spike' \citep{Nather1981,Marsh1999}. 
The central spike is redshifted by an amount approximately consistent with a gravitational potential expected on or near the white dwarf surface. 
Wavelength-dependent blue-shifts are also seen in the central spike of some \hei\ lines, resulting from forbidden transitions excited in regions of high electron density \citep{Morales-Rueda2003}. 
On top of these systemic velocity offsets relative to the rest wavelength, the feature undergoes low-amplitude velocity changes which appear to coincide with the phase and amplitude of the central white dwarf \citep{Kupfer2016}.
The central spike has therefore been assumed to originate at the central white dwarf, and these phase-dependent Doppler shifts have regularly been used to measure AM\,CVn mass ratios. 
However, there has been no direct test of this assumption.

The best-characterised AM\,CVn is Gaia14aae, the only known system in which the central white dwarf is fully eclipsed \citep{Campbell2015}. Its orbital period of 49.7\,minutes puts Gaia14aae at the long-period end of the AM\,CVn orbital distribution.
Modelling the eclipse lightcurves of Gaia14aae has allowed the component stellar masses and radii to be tightly constrained \citep{Green2018}. The properties measured in that work are summarised in Table~\ref{tab:av-results}. 
An estimate of the mass transfer rate based on the \textit{Gaia} parallax of the system was consistent with these stellar masses \citep{Ramsay2018}.
The tight constraints on Gaia14aae make it an ideal system to test the assumptions made for other AM\,CVns.

In this paper, we present and analyse phase-resolved spectroscopy of Gaia14aae. In Section~\ref{sec:observations} we describe the observations undertaken for this project. In Section~\ref{sec:results} we present the spectra of Gaia14aae and their analysis. In Section~\ref{sec:discussion} we discuss the implications of these findings for Gaia14aae in the context of its formation history and the population of AM\,CVns.

\begin{table}
	\centering
	\caption{Summary of the properties of Gaia14aae based on eclipse photometry, as presented in \citet{Green2018}. \review{$1 \sigma$} uncertainties are quoted in brackets. We additionally predict the projected orbital velocity, $K_\mathrm{WD}$, of the accretor, based on stellar masses, orbital separation, and orbital inclination.
	}
	\label{tab:av-results}
	\begin{tabular}{lc} 
		\hline
		Parameter & Value\\
		\hline
		Mass ratio, $q$ & $0.0287(20)$\\
		Inclination, $i$ ($^\circ$) & $86.3(3)$\\
		Orbital separation, $a$ ($R_\odot$) & $0.430(3)$\\
		Accretor mass, $M_1$ ($M_\odot$) & $0.87(2)$\\
		Donor mass, $M_2$ ($M_\odot$) & $0.0250(13)$\\
		Accretor radius, $R_1$ ($R_\odot$) & $0.0092(3)$\\
		Donor radius, $R_2$ ($R_\odot$) & $0.060(10)$\\
		Disc radius, $R_\text{disc}$ ($R_\odot$) & $0.264(3)$\\
		Bright spot separation, $R_\text{spot}$ ($R_\odot$) & $0.187(2)$\\
		\hline
		Accretor orbital velocity, $K_\mathrm{WD}$ (km/s) & $17.6 \pm 1.0$\\
		\hline
	\end{tabular}
\end{table}


\section{Observations}
\label{sec:observations}

Phase-resolved spectroscopy of Gaia14aae was obtained using the ISIS spectrograph on the 4.2\,m William Herschel Telescope (WHT), and the OSIRIS spectrograph \citep{Sanchez2012} on the 10.4\,m Gran Telescopio Canarias (GTC), both at the Roque de los Muchachos observatory on La Palma. The observations undertaken for this project are summarised in Table~\ref{tab:observations}. Conditions for all observations were clear with good seeing (0.4--1").
All images were bias-subtracted and flat field-corrected using tungsten lamps.
Extraction and calibration were carried out with optimal weights, using the software packages \texttt{\sc Pamela}\footnote{github.com/starlink/starlink/tree/master/applications/pamela} and \texttt{\sc Molly}\footnote{deneb.astro.warwick.ac.uk/phsaap/software/}.

The WHT spectra were collected using a 1" slit and exposure times of 240s. The ISIS spectrograph has two arms, allowing it to simultaneously collect data at blue and red wavelengths. 
In the blue arm we used the R300B grating, giving a mean dispersion of 1.7\,\AA/pixel across wavelengths 3500--5400\,\AA. In the red arm the R316R grating and a GG495 order-sorter filter were used, giving a mean dispersion of 1.8\,\AA/pixel across wavelengths 5400--8000\,\AA. The resolutions were 3.5\,\AA\ in the blue arm and 3.2\,\AA\ in the red arm, both measured using arc lines close to the centre of the spectral range.

Wavelength calibration arcs were obtained once per hour throughout the night. 
The red arm was calibrated against 32 arc lines using a five-term polynomial, achieving a root-mean-square (RMS) deviation from the polynomial of 0.017\,\AA. 
The blue arm was calibrated against 30 arc lines using a seven-term polynomial, achieving an RMS deviation of 0.175\,\AA.
Due to a lack of viable arc lines at the blue end of the spectrum, the calibration for wavelengths $3500$--$3900$\,\AA\ is unreliable.
Flux calibration was carried out against the standard star BD+33~2642, observed on both nights.

The GTC spectra were collected using a 0.8" slit and exposure times of 70s. The R2500V grating was used, giving a dispersion 0.8\,\AA/pixel across wavelengths 4400--6000\,\AA. The resolution was 2.1\,\AA\ based on a fit to the arc lines, or 2.3\,\AA\ based on the 5577\,\AA\ sky emission line. 
Wavelength calibration was carried out against 27 arc lines using a five-term polynomial. The resulting RMS was 0.016\,\AA. The flux was calibrated against the standard star Feige~66, observed on the same night.

\begin{table}
	\centering
	\caption{Observations undertaken for this work.}
	\label{tab:observations}
	\begin{tabular}{lcc}
		\hline
		Date & Instrument & Exposures\\
		\hline
		2015 July 16 & WHT+ISIS & 31$\times$240\,s\\
		2015 July 17 & WHT+ISIS & 51$\times$240\,s\\
		2016 May 14 & GTC+OSIRIS & 136$\times$70\,s\\
		\hline
	\end{tabular}
\end{table}


\section{Results}
\label{sec:results}

\subsection{Average Spectrum}
\label{sec:avspec}

\begin{figure*}
	\includegraphics[width=500pt]{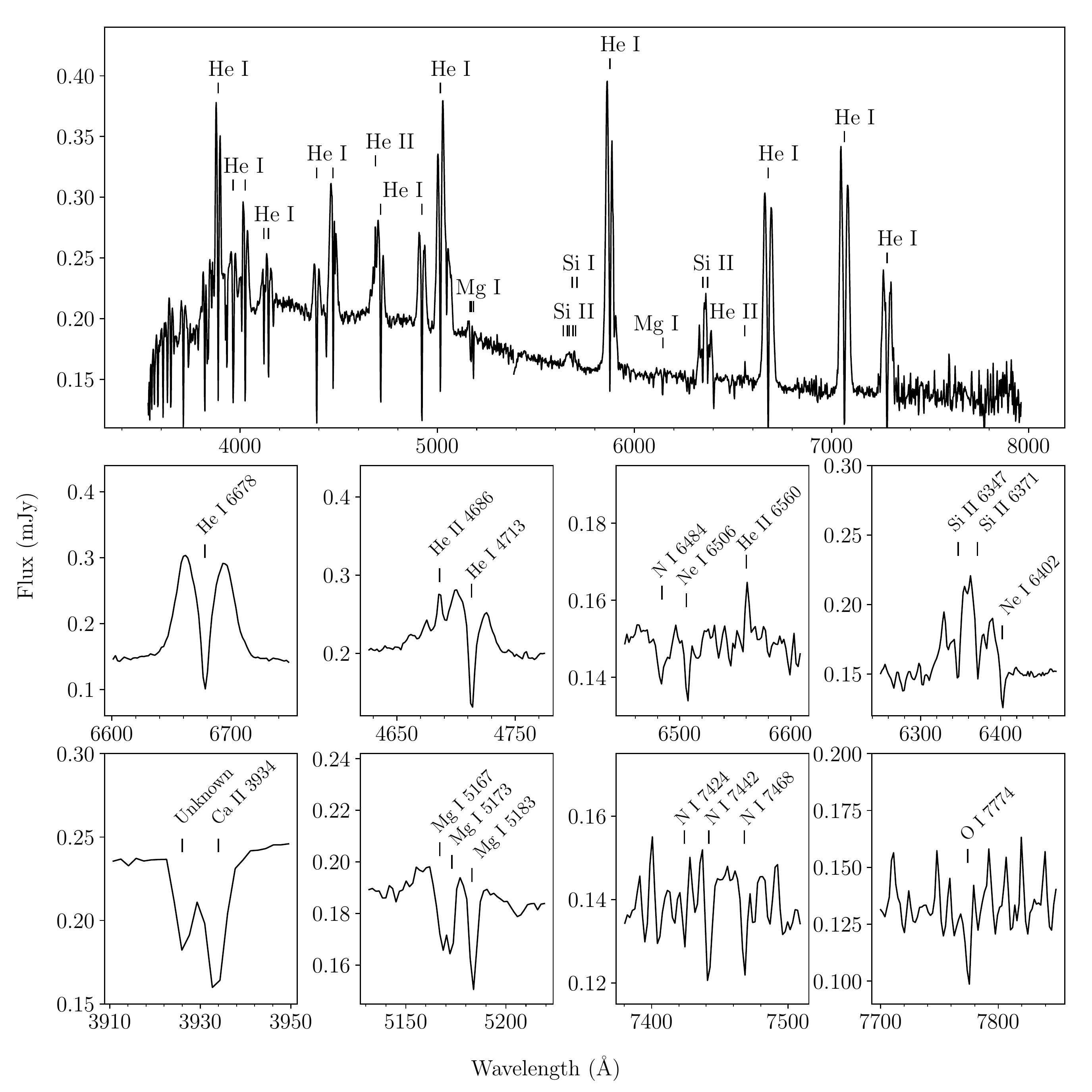}
    \caption{\textit{Above:} The average WHT spectrum, obtained in 2015 July, created by combining 78 spectra with 240s exposure times that did not coincide with eclipses. The join between the red and blue arms of the spectrum is around 5400\,\AA.
     The positions of various identified spectral features are marked.
     For wavelengths $< 3900$\,\AA, the wavelength calibration is unreliable and we therefore do not identify these features.
      \textit{Below:} Close-ups of selected spectral features. As can be seen, detections of \ion{N}{I} and \ion{O}{I} are marginal relative to the noise. 
      Several absorption lines are still unidentified, including the feature next to the \ion{Ca}{II} \textit{K} line.
    }
    \label{fig:wht-spec-g14}
\end{figure*}

\begin{table}
	\centering
	\caption{Features identified from WHT and GTC spectra of Gaia14aae, along with their measured EWs. The EW of some features could not be measured, due either to a blend or insufficient signal. The quoted uncertainties are scaled to account for any scatter in the local continuum. Regardless, we note that the uncertainties are statistical only, and do not represent the false alarm probability of a particular line. We class the \hei\ lines as emission features, but note that these features also have prominent absorption components which in some cases result in positive EWs.
	}
	\label{tab:ews}
	\begin{tabular}{lcc} 
		\hline
		Line (wavelength in \AA) & WHT EW (\AA) & GTC EW (\AA) \\
		\hline
\textit{Emission features} &&\\
\hei\ 3888.6 & $-9.2 \pm 0.5$ & -- \\
\hei\ 3964.7 & $1.45 \pm 0.28$ & --\\
\hei\ 4026.2 & $-4.39 \pm 0.27$ & --\\
\hei\ 4120.8/43.8 & $-3.25 \pm 0.27$ & -- \\
\hei\ 4387.9 & $-2.09 \pm 0.16$ & --\\
\hei\ 4471.5 $^*$ & $-13.42 \pm 0.31$ & $-7.39 \pm 0.26$\\
\heii\ 4685.7  $^*$ & X$^{a}$ & X$^{a}$ \\
\hei\ 4713.1 $^*$ & $-12.37 \pm 0.32$ & $-6.30 \pm 0.22$\\
\hei\ 4921.9 & $-9.72 \pm 0.19$ & $-4.61 \pm 0.14$\\ 
\hei\ 5015.6/47.7 & $-35.8 \pm 0.5$ & $-17.77 \pm 0.32$\\
\siii\ 5640--5710 $^\dagger$ & $-3.19 \pm 0.15$ & $-1.30 \pm 0.16$\\
\hei\ 5875.6 $^*$ & $-46.4 \pm 1.0$ & $-33.2 \pm 0.9$\\
\siii\ 5957.6/78.9 $^*$ & $-0.38 \pm 0.11$ & $-0.49 \pm 0.16$ \\
\siii\ 6347.1/71.4 & $-15.7 \pm 0.5$ & --\\
\heii\ 6560.1 & $-0.82 \pm 0.16$ & --\\
\hei\ 6678.2 & $-42.1 \pm 1.2$ & --\\
\hei\ 7065.2 & $-51.4 \pm 1.5$ & --\\
\hei\ 7281.4 & $-26.7 \pm 1.2$ & --\\



&&\\
\textit{Absorption features} &&\\
\ion{Mg}{I} 3829.4/32.3/38.3 & $4.47 \pm 0.08$ & --\\
\siii\ 3856.0 $^*$ & X & --\\
\ion{Ca}{II} 3933.7 $^{* \ddagger}$ & X & --\\
\ion{Mg}{II} 4481.2 $^*$ & X & X \\
\ion{Mg}{I} 5167.3/72.7/83.6 & $0.82 \pm 0.12$ & $1.38 \pm 0.11$ \\
\ion{Na}{I} 5890.0 $^*$ & X & X \\
\ion{Na}{I} 5895.2 $^*$ & X & X \\
\ion{Mg}{I} 6145.1 & $0.53 \pm 0.09$ & --\\
\ion{Ne}{I} 6402.2 $^*$ & X  & --\\
\ion{N}{I} 6483.7 & $0.67 \pm 0.08$ & --\\
\ion{Ne}{I} 6506.5 & $0.58 \pm 0.08$ & --\\
\ion{N}{I} 7423.6/42.3/68.3 & X & --\\
\ion{O}{I} 7774.2 & $1.5 \pm 0.5$ & --\\

		\hline
\multicolumn{3}{l}{--\,\, Wavelength not covered.}\\
\multicolumn{3}{l}{$^*$\,\,\, Marked features are blended with their neighbours.
}\\
\multicolumn{3}{l}{$^\dagger$\,\,\, Blended feature consistent with several Si\,II emission lines.}\\
\multicolumn{3}{l}{$^\ddagger$\,\,\, \ion{Ca}{II} 3968\,\AA\ was excluded due to a blend with \hei\ 3965\,\AA.}\\
\multicolumn{3}{l}{$^{a}$\,\,\, \review{EW for \hei\ 4713\,\AA\ includes blended \heii\ 4686\,\AA\ line.}}\\
\multicolumn{3}{l}{X\, Line present but EW could not be measured, \review{due to either}}\\
\multicolumn{3}{l}{\,\,\,\,\,\, \review{a blend with another line or significant continuum noise.}}\\
\hline
	\end{tabular}
\end{table}

\subsubsection{Helium lines}

The average spectrum of Gaia14aae is shown in Figure~\ref{fig:wht-spec-g14}. It consists of a blue continuum with a series of strong helium emission lines, and some weaker metal lines which are in a mixture of emission and absorption. The spectral lines identified and their measured equivalent widths (EWs) are listed in Table~\ref{tab:ews}.
The majority of these spectral lines originate from \hei, and have a double-peaked profile characteristic of accretion disc emission, along with deep absorption cores. 
Also clearly visible is an emission line from \heii\ at 4685.7\,\AA. Unlike the \hei\ emission lines, \heii\ appears single-peaked in the average spectrum. We identify this emission as the `central spike' feature.
There is no evidence for the central spike feature in any \hei\ line.

A weak emission line is seen at 6560\,\AA\ which may originate from the H$\alpha$ line (6562.8\,\AA) or from a line in the \heii\ Pickering series (6560.1\,\AA). As discussed in Section~\ref{sec:hydrogen}, we believe this to be \heii.

Absorption cores of the type seen in Gaia14aae are commonly seen in CV and AM CVn systems when the system is viewed edge-on relative to the orbital plane. These cores result from obscuration of the central white dwarf by an accretion disc which is optically thick at those wavelengths and optically thin elsewhere. The narrow profile of the absorption cores in Gaia14aae suggests that they are dominated by this form of absorption rather than photospheric absorption, which, when seen in helium-atmosphere white dwarfs, typically produces broader spectral lines \citep{Bergeron2011}. We discuss this further in Section~\ref{sec:eclipse}.

\begin{figure}
	\includegraphics[width=\columnwidth]{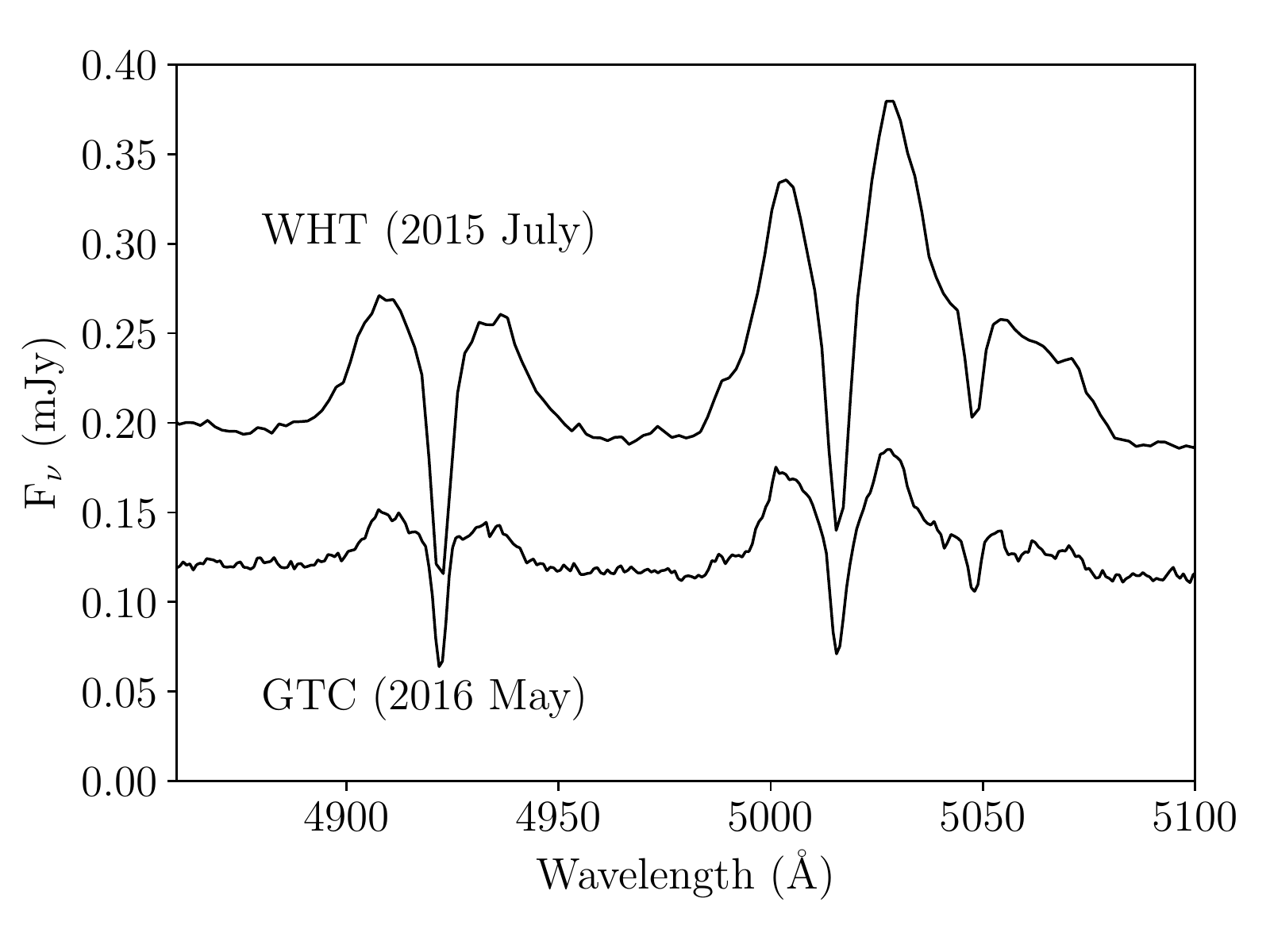}
    \caption{Comparison of WHT and GTC average spectra for a selected wavelength range. No flux offsets or flux scaling have been applied, apart from the original flux calibration described in Section~\ref{sec:observations}. 
    The difference in continuum level may be genuine or may be a result of slit losses. 
    However, there is a visible decrease in the strength of the emission lines relative to the continuum by the time of the GTC spectrum.
    }
    \label{fig:spec_zoom}
\end{figure}

Between the WHT observations in 2015 July and the GTC observations in 2016 May, the EW of He\,I lines decreased by a mean factor of 0.65. 
Gaia14aae underwent an outburst in 2014 \citep{Campbell2015} and, at the time of these observations, was still decreasing in brightness towards quiescence \citep{Green2018}. 
A comparison of WHT and GTC average spectra is shown in Fig.~\ref{fig:spec_zoom}. There is a clear decrease in both the continuum level and the line strengths by the time of the GTC spectra in 2016 May. This may be due to slit losses or, if intrinsic to the system, may be a result of the system continuing to cool following its outburst. The decrease in EW may be explained by the line strengths decreasing more significantly than the continuum level and the absorption cores.
Similar changes in EW have been seen in dwarf novae as the system cools following an outburst \citep[eg.\ Fig.~1 of][]{Szkody2012}.

The measured \hei\ EWs can be compared with other AM\,CVns. For a sample of 23 AM\,CVns with emission line spectra, \citet[their Fig.~10]{Carter2013} found a correlation between orbital period and the EW of the \hei\ 5875\,\AA\ line. Based on this correlation, the 5875\,\AA\ line of Gaia14aae is weaker than expected, though within the scatter. This relative weakness is likely due to the absorption cores that result from the edge-on viewing angle of Gaia14aae.
We note, however, that the partially-eclipsing AM\,CVn binary YZ\,LMi (which is necessarily viewed edge-on) does not show such strong absorption cores \citep{Anderson2005}.

\subsubsection{Metal lines}

In Gaia14aae we detect metal lines from a variety of elements listed in Table~\ref{tab:ews}, including N, O, Ne, Na, Mg, Si, and Ca.
Of these, the only prominent \textit{emission} lines are \siii\ lines at 6347 and 6371\,\AA, along with a blended \siii\ feature around 5640--5710\,\AA. \citet{Marsh1991} predicted that \siii\ 6347 and 6371\,\AA\ should be among the strongest metal emission lines in helium-dominated, optically thin accretion discs. 

All other metal features in the spectrum of Gaia14aae are absorption lines.
Metal absorption lines are relatively rare in quiescent AM\,CVn binaries, though they have been previously detected in SDSS\,J1552+3201 \citep{Roelofs2007b}, SDSS\,J1208+3550, and SDSS\,J1642+1934 \citep{Kupfer2013}. All three of these systems show \ion{Mg}{I} absorption, and the latter two additionally show \ion{Si}{I/II} absorption. 
\ion{N}{I} and \ion{N}{II} absorption has been seen in GP\,Com and SDSS\,1908+3940 \citep{Morales-Rueda2003,Kupfer2015,Kupfer2016}, with GP\,Com also showing significant N emission lines.
O and Ne have not been previously detected in absorption, but are present in emission in GP\,Com and V396\,Hya \citep{Kupfer2016}.
Similarly, Ca has previously been detected in emission in systems such as V406\,Hya \citep{Roelofs2006b}, but never before in absorption.
The \ion{Ca}{II} \textit{K} line is seen in our spectrum, while the nearby \ion{Ca}{II} \textit{H} line at 3968\, \AA\ would be blended with a \hei\ line at 3965\,\AA\ and is therefore not detected.
The detection of Na appears to be unprecedented in AM\,CVn systems. However, we note that the Na absorption lines in our spectra may be residuals of our sky subtraction, as the Na emission from the sky was strong on the nights of our observations.

The detection of a wide variety of metals in absorption is unusual in an AM\,CVn, but not necessarily unexpected for Gaia14aae. Under the assumption that these absorption lines result from obscuration of the central white dwarf, the high inclination of Gaia14aae is likely to make such otherwise weak absorption features more prominent. 

In Table~\ref{tab:ews} we have not identified any lines of Fe. This is something of a surprise; Fe is a common element and has been detected in a number of AM\,CVns \citep[eg.][]{Roelofs2005a,Roelofs2006b,Roelofs2007a,Kupfer2013}. 
Alongside \siii, \citet{Marsh1991} predicted that \ion{Fe}{II} at 5169\,\AA\ should be among the strongest metal emission lines.
The presence of \siii\ and absence of \ion{Fe}{II} is consistent with some previously studied AM\,CVn systems, while others have shown both \siii\ and \ion{Fe}{II} emission or no detectable metal emission \citep[eg.][]{Kupfer2013,Kupfer2016}.
We note that we cannot rule out that some spectral features may result from Fe. 
The absorption series in the 3500--3900\,\AA\ range (which we were unable to precisely wavelength calibrate) is most likely to be \hei, but some lines may be explained by Fe.
The absorption lines that we identify as \ion{Mg}{I} 5167 and 5173\,\AA\ may in fact be \ion{Fe}{I} 5167 and 5172\,\AA, although there is no Fe line corresponding to \ion{Mg}{I} 5183\,\AA.
The unknown feature around 3927\,\AA\ (Fig.~\ref{fig:wht-spec-g14}, lower left panel) could be \ion{Fe}{I} 3928\,\AA, though it would have to be blueshifted where most features are slightly redshifted, and we do not see the nearby \ion{Fe}{I} transitions at 3923 and 3930\,\AA\ despite their similar predicted strength.
We also note non-detections of \ion{Fe}{I} at 4064, 4272, 4282, 4308 4325, 4958, 5227, and 5270\,\AA, and of \ion{Fe}{II} at 6248 and 6456\,\AA, all of which are expected to be reasonably strong and occur at regions of clear baseline in our spectrum.

The presence of N and O is of particular interest, as the abundances of CNO elements can give insights into the evolutionary history of the donor and hence of the binary \citep{Yungelson2008, Nelemans2010}. If the donor has been through a phase of helium burning, C and O will be significantly enriched relative to N. 
We do not detect C in Gaia14aae, but given the marginal nature of the detections of N and O this is not a significant non-detection.

\subsubsection{Radial Velocity}
\label{sec:rv}

\begin{figure}
	\includegraphics[width=\columnwidth]{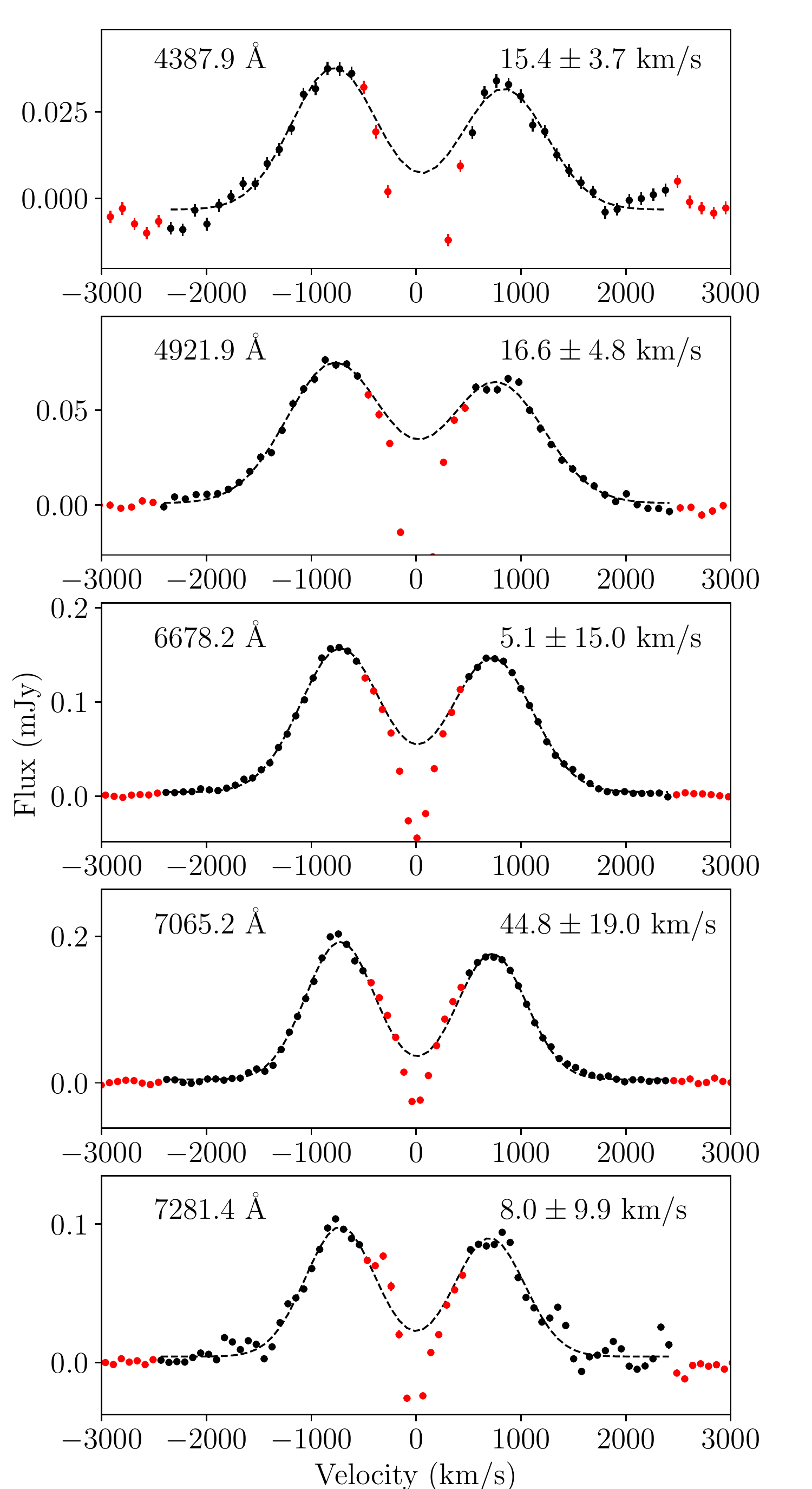}
    \caption{Emission lines of \hei\ with best-fitting double-peaked Gaussian profiles. Lines which are blended with neighbours or strongly asymmetric were omitted from the fitting process. The central absorption region of each line was masked (masked points are indicated in red). The measured systemic velocity of each line is quoted in the top right of the panel.
    }
    \label{fig:rv-fit-g14}
\end{figure}

\begin{figure}
	\includegraphics[width=\columnwidth]{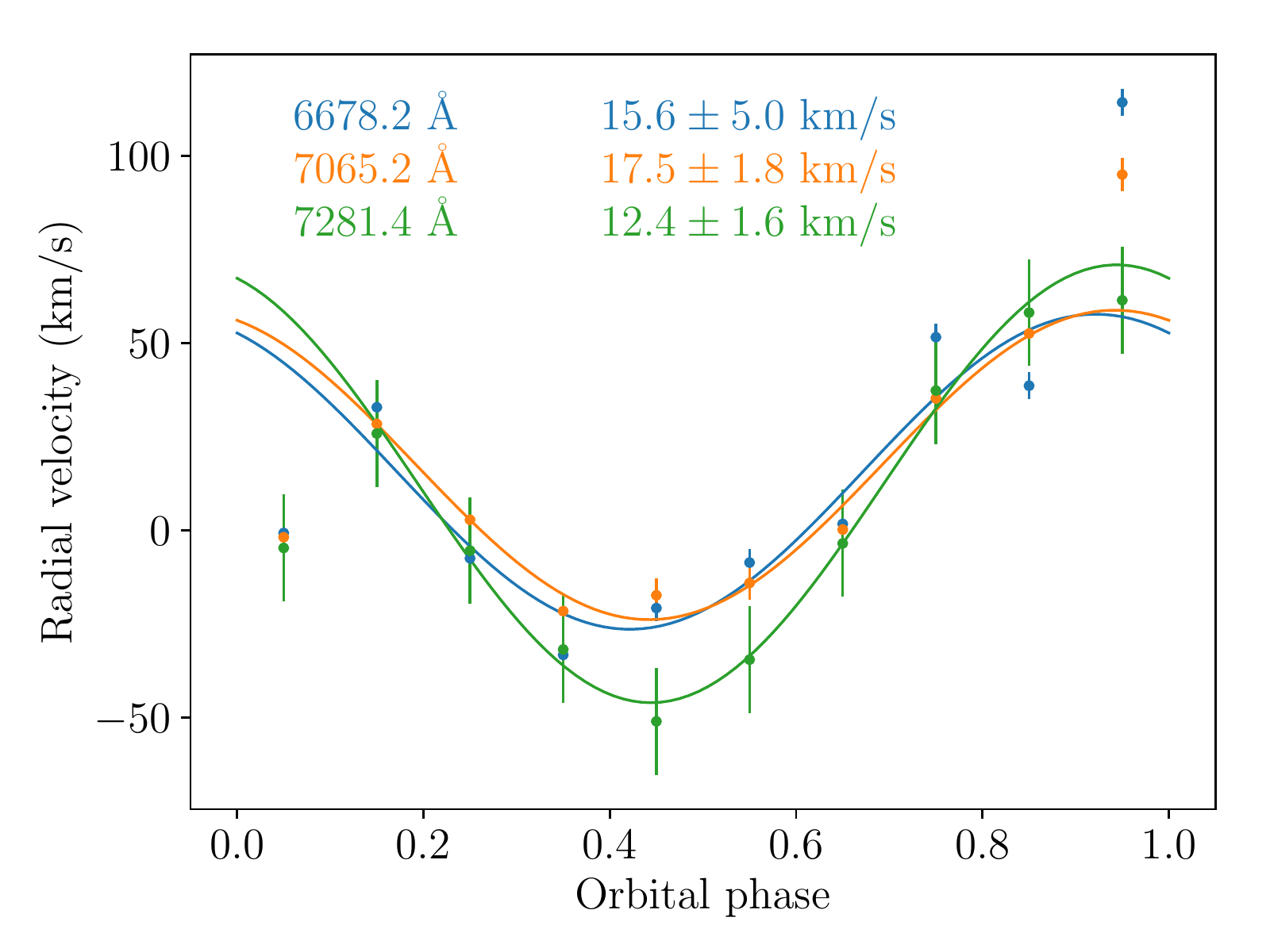}
    \caption{Radial velocities measured from phase-binned and averaged WHT spectra for three \hei\ lines. Radial velocity measurements are colour-coded according to the spectral line to which they correspond. Sinusoids were fitted to measurements from each spectral line, and are also shown.
    Note that the phase bins centred on phases 0.05 and 0.95 are affected by the eclipse of the disc, and were therefore not considered during the sinusoidal fit.
    }
    \label{fig:rv-sine-g14}
\end{figure}

We measured the average radial velocity of Gaia14aae from the \hei\ emission lines in our WHT spectra \review{in several different ways}. First, we selected \hei\ lines that were not blended with other lines, and avoided lines that were extremely asymmetric. This left 5 suitable \hei\ lines, shown in Fig.~\ref{fig:rv-fit-g14}. 
For measuring the radial velocity we are most interested in the outer edges of the line profile. The outer edges correspond to the inner edge of the accretion disc and are therefore less likely to be affected by asymmetries in the disc that may occur due to the bright spots or eccentricity of the disc. In Gaia14aae the centre of each line is also affected by the absorption cores. We therefore masked velocities $< 500$\,km/s in each line.
A double-peaked Gaussian fit to these lines was performed, with the widths of the Gaussians held equal but their heights allowed to vary, from which the systemic radial velocity, $\gamma$, was found as the midpoint of the centres of the two Gaussians. 
This process found radial velocities consistent with each other, though some have large error bars. Their average systemic velocity $\gamma = 15.5 \pm 2.7$\,km/s.

\review{
We note the possibility for the mean spectrum to be biased if the sampled spectra are not uniformly distributed across orbital phase. 
We therefore produced ten phase-binned and averaged spectra.
The radial velocity of each spectrum was measured using a double-Gaussian fit as before.
In Fig.~\ref{fig:rv-sine-g14} we show sinusoidal fits to these radial velocity measurements.
The 4388\,\AA\ and 4921\,\AA\ lines showed a significant number of anomalous radial velocity measurements and were excluded (though we note that, due to the large uncertainties on the sinusoidal fits to these lines, including or excluding these lines does not significantly change the weighted mean).
The weighted mean $\gamma$ from these fits was found to be $14.7 \pm 1.2$\,km/s, consistent with the above.
}

\review{
In Section \ref{sec:doppler} we perform a sinusoidal fit to the radial velocity of the bright spot of \heii~4686\,\AA, which is found to be $16.0 \pm 5.7$\,km/s, which is again consistent with the values discussed previously.
}

\review{
However, we note that there are several sources of uncertainty in $\gamma$ which these methods do not counteract.
Phase-dependent variations in the bright spot flux are possible, for instance due to partial obscuration of the bright spot by the disc, and could potentially distort measured radial velocities. 
The AM\,CVn binary GP\,Com shows measurably non-sinusoidal variations in the radial velocity of its bright spot \citep{Marsh1999}; if the radial velocity curve of the bright spot in Gaia14aae is similarly non-sinusoidal, this may again bias our measurements of $\gamma$ which are based on sinusoidal fits.
}

\subsubsection{Upper Limit on Hydrogen}
\label{sec:hydrogen}

The spectral feature near 6560\,\AA\ has two possible origins, a \heii\ line (6560.1\,\AA) or H$\alpha$ (6562.8\,\AA). We measured the radial velocity shift required to produce this feature in each interpretation, assuming a systemic redshift $\gamma$ of 16\,km/s to match that measured in Sections~\ref{sec:rv} and \ref{sec:doppler}. In the \heii\ case, a redshift of $46 \pm 17$\,km/s is required, which agrees well with the redshift measured for the strong \heii\ 4686\,\AA\ central spike (Section~\ref{sec:doppler}). In the H$\alpha$ case, a blue-shift of $-78 \pm 18$\,km/s would be required. As we do not know of any mechanism that would produce such a blue-shift, we feel confident in identifying this feature as \heii\ emission.

As the hydrogen content of Gaia14aae is of interest regarding its evolutionary history, we attempted to measure upper limits on the EW of H$\alpha$ emission. We did this for two assumed line profiles of H$\alpha$: firstly, a single-peaked Gaussian with the same width and the same baseline redshift as the \heii\ 6560\,\AA\ line, and secondly a double-peaked Gaussian profile with properties held at the average of the \hei\ velocity profiles shown in Fig.~\ref{fig:rv-fit-g14}. For this process, the \heii\ 6560\,\AA\ was first subtracted from the spectrum using a single Gaussian fit. The uncertainties on each data point in the spectrum were then scaled such that they were consistent with the scatter on data points in the region around 6562\,\AA. Assuming Gaussian uncertainties, we derived 3\,$\sigma$ limits on the H$\alpha$ EW of -0.44\,\AA\ in the single-peaked case, and -1.14\,\AA\ in the double-peaked case.

Quantifying an upper limit on the hydrogen fraction in the accreted material would require modelling of the accretion disc. Such modelling has been done for several AM\,CVn binaries, most recently non-LTE modelling of V396\,Hya \citep[aka CE\,315,][]{Nagel2009}. 
As Gaia14aae and V396\,Hya are similar in some ways (most importantly orbital period and mass transfer rate), we compare our upper limits to the predicted spectra of that work \citep[their Fig. 9]{Nagel2009}.
Based on this comparison, we cannot rule out a hydrogen fraction of $10^{-5}$ (predicted EW $\approx$ 0.6\,\AA), and it seems unlikely we would be able to rule out $10^{-4}$ either. 
This represents an easing of the upper limit of $10^{-5}$ assumed in \citet{Green2018}.


\subsection{In-Eclipse Spectrum and White Dwarf Spectrum}
\label{sec:eclipse}

\begin{figure*}
	\includegraphics[width=500pt]{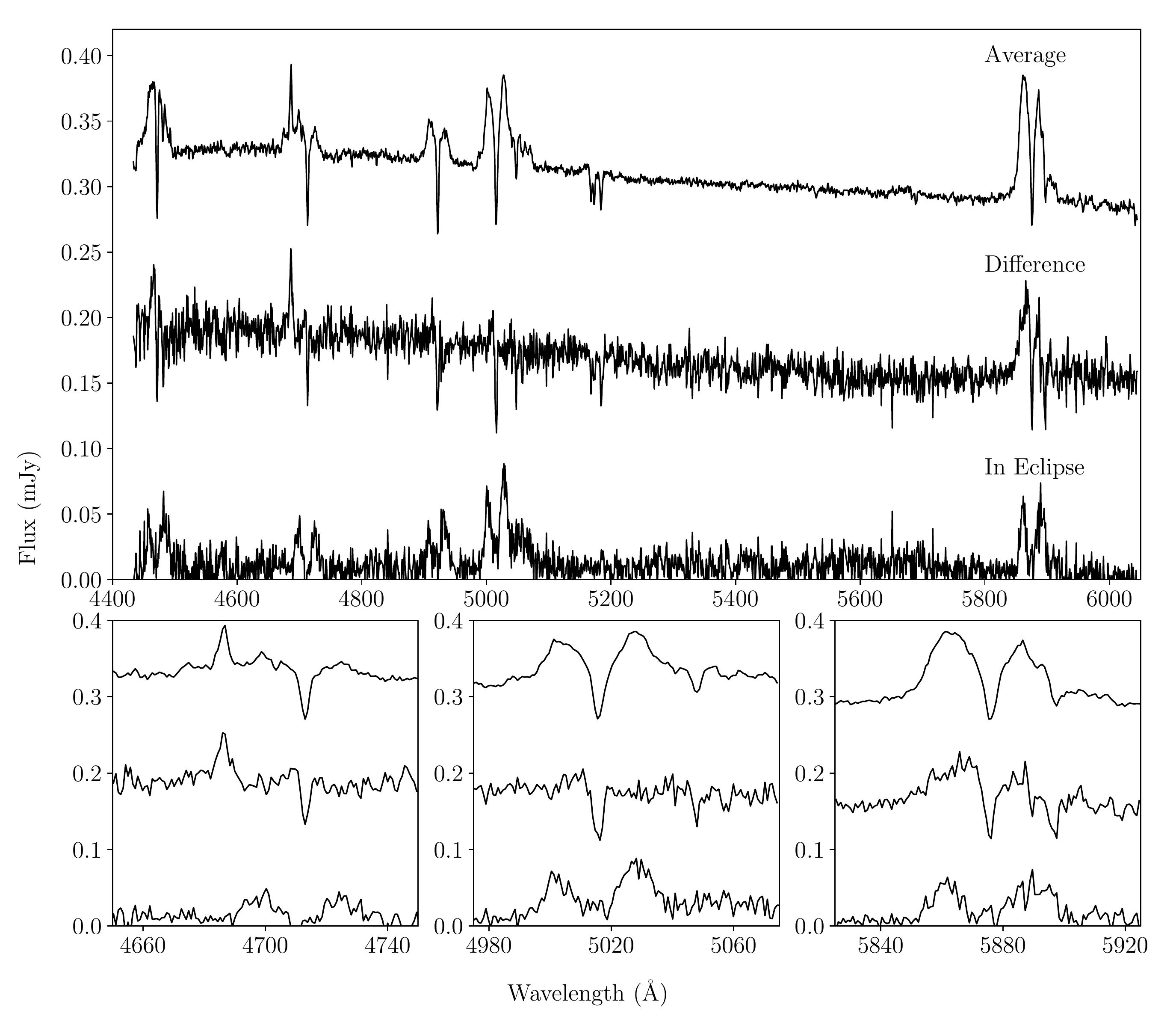}
    \caption{\textit{Above:} GTC spectra of Gaia14aae. The top line shows the average of all out-of-eclipse spectra, offset by +0.2\,mJy. The bottom line shows the spectrum taken during eclipse. The middle line shows the difference between the two spectra, offset by +0.07\,mJy. 
    The difference spectrum combines the spectra of the white dwarf and of a region of the inner disc that is eclipsed at the same time. It is characterised by \hei\ absorption lines (the cores of the \hei\ features seen in the average spectrum) and \heii\ emission at 4686\,\AA. Most \hei\ features show little or no emission in the eclipse spectrum, though some emission is seen at the \hei\ 5875\,\AA\ line. 
    \textit{Below:} Close-ups of selected \hei\ and \heii\ features: 4686 and 4713\,\AA\ (left), 5015 and 5048\,\AA\ (centre), and 5875\,\AA\ (right).
    }
    \label{fig:eclipse-spec-g14}
\end{figure*}

\begin{figure*}
	\includegraphics[width=500pt]{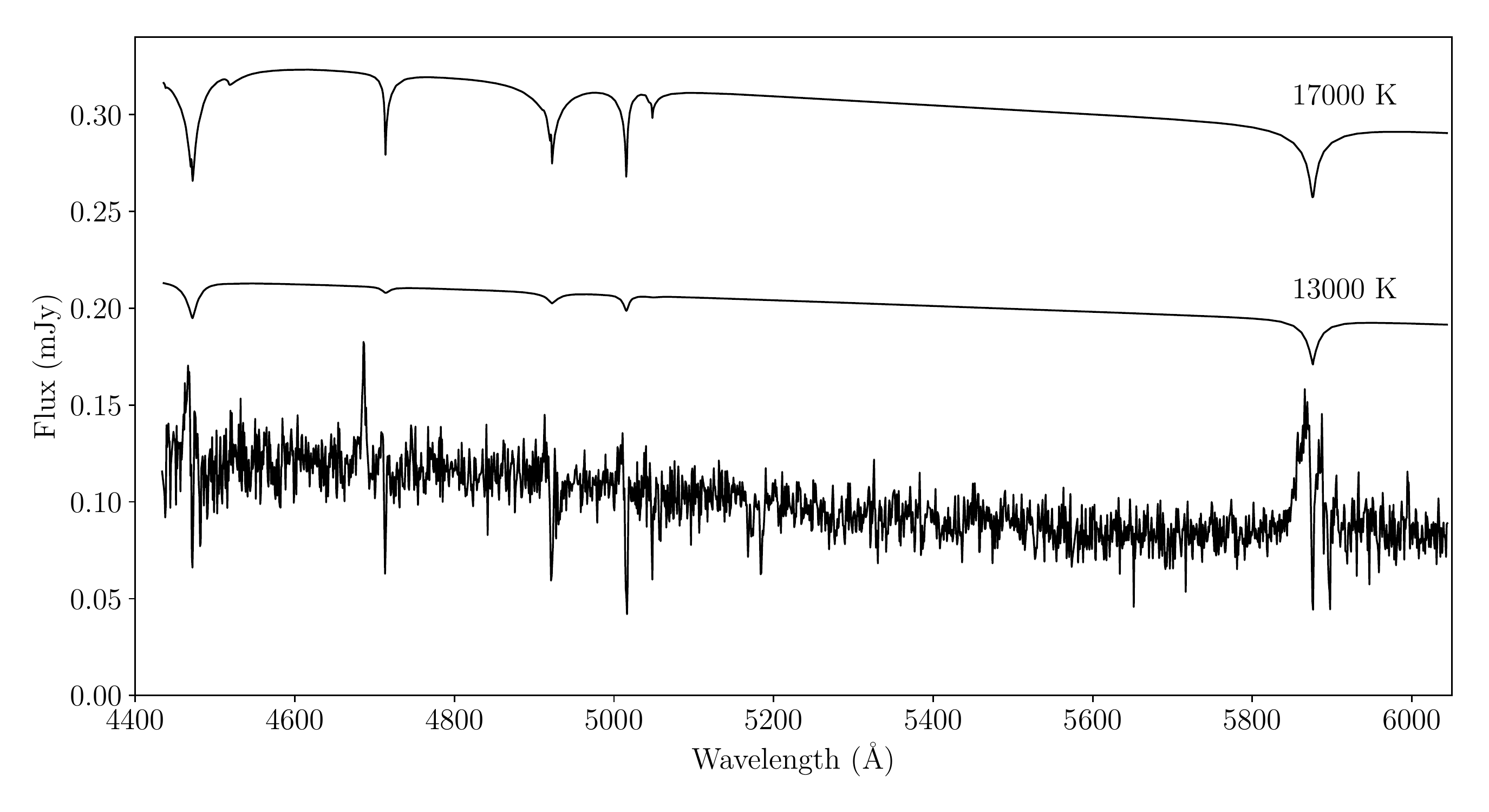}
    \caption{Spectrum of the eclipsed light of Gaia14aae from Fig.~\ref{fig:eclipse-spec-g14}, compared with model spectra for pure helium atmosphere white dwarfs at the two proposed temperatures for the central white dwarf, 13000\,K and 17000\,K. The 13000\,K and 17000\,K models were scaled such that the continuum matched the continuum of Gaia14aae, then offset by +0.1 and +0.2\,mJy respectively. It is clear that the absorption lines in the observed spectrum are narrower and deeper than the lines in either model, suggesting a significant contribution from absorption of the white dwarf light by the accretion disc.
    }
    \label{fig:model-spec-g14}
\end{figure*}

As the first known fully-eclipsing AM\,CVn binary, Gaia14aae presents the opportunity to disentangle the \review{spatial} components of the system. Because the eclipse duration is slightly longer than the exposure time of our GTC observations (70\,s), we were able to obtain one spectrum contained entirely within the eclipse of the central white dwarf. The difference between this spectrum and the out-of-eclipse average spectrum then gives the spectrum of the eclipsed region of Gaia14aae, a region which includes the central white dwarf and part of the inner disc which is eclipsed simultaneously. In Fig.~\ref{fig:eclipse-spec-g14} we show the spectrum during eclipse, the average spectrum outside of eclipse, and the difference spectrum (the spectrum of the eclipsed light).

The strong central spike emission feature from the \heii\ 4686\,\AA\ line is clearly visible in the difference spectrum in  Figure~\ref{fig:eclipse-spec-g14}, and is not visible at all during eclipse. 
This feature must therefore arise from a region which is occulted by the donor at this phase. Likely suggestions are the central white dwarf itself, a region near the central white dwarf, or the inner accretion disc.

\review{The majority of \hei\ emission is still visible during eclipse, as expected because at any one time the majority of the disc is still visible.}
However, some amount of \hei\ emission is seen in the difference spectrum in Figure~\ref{fig:eclipse-spec-g14}, most clearly around 5875\,\AA. There are two possible sources for this emission: this may be emission from the regions of the disc which are eclipsed, especially the inner disc; or it may be a residual artefact of the emission in the average spectrum.

The absorption cores of \hei\ lines are visible in the difference spectrum of Figure~\ref{fig:eclipse-spec-g14} as deep, sharp absorption lines. 
As was mentioned in Section~\ref{sec:avspec}, these lines are narrower than would be expected from photospheric absorption. In Figure~\ref{fig:model-spec-g14} we compare the difference spectrum to synthetic spectra of 17000K and 13000K white dwarfs, both with log($g$)=8.5 and pure-helium (DB spectroscopic class) atmospheres. 
These models are identical to those presented by \citet{Bergeron2011}.

The absorption seen from the white dwarf in Gaia14aae is both narrower and deeper than either of these model spectra. 
This is consistent with an understanding that the disc is less transparent at the frequencies of these lines than it is in the continuum.
If the accretion disc scale height is negligible compared to the white dwarf radius, as is expected, then slightly less than 50\% of the white dwarf light will pass through the accretion disc. 
For some \hei\ lines, in particular at 5015 and 5875\,\AA, the depth of the absorption core is over 50\% of the continuum, therefore requiring a contribution from photospheric absorption as well.

\subsection{Trailed Spectra and Doppler Maps}
\label{sec:doppler}

\begin{figure*}
	\includegraphics[width=500pt]{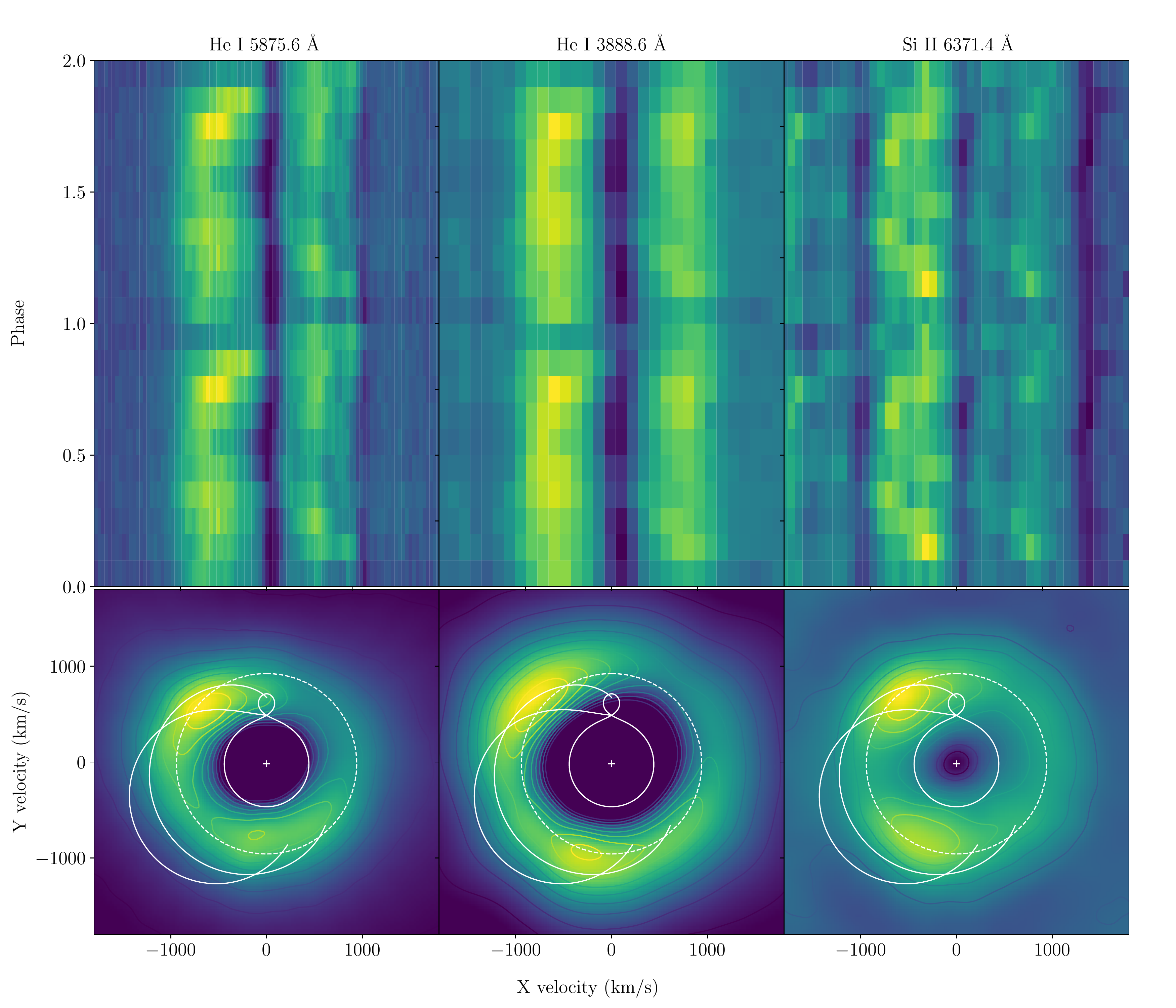}
    \caption{\textit{Above:} Phase-folded, trailed spectra of \hei\ and \siii\ lines using our data from the GTC (left panel) and WHT (centre and right panels). Absorption is shown in blue (dark for greyscale prints), and emission in yellow (light). The features show central absorption, a strong S-wave (maximum blue excursion at phase 0.7), and disc emission (either side of the central absorption). 
    \textit{Below:} Doppler maps produced from the individual spectra. In addition to the central absorption and disc emission, two bright spots can be seen. The second spot is weaker than the first, though by how much varies depending on the emission line.
    On the Doppler maps we also plot the predicted velocities of the Roche lobes (solid lines), the central white dwarf (cross), the stream of infalling matter (solid lines), and the velocity corresponding to the photometrically-measured bright spot separation radius (dotted line). The rotation of the map is such that the predicted position of the central white dwarf is below the origin of the map.
    }
    \label{fig:trail-wht-g14}
\end{figure*}

\begin{figure*}
	\includegraphics[width=500pt]{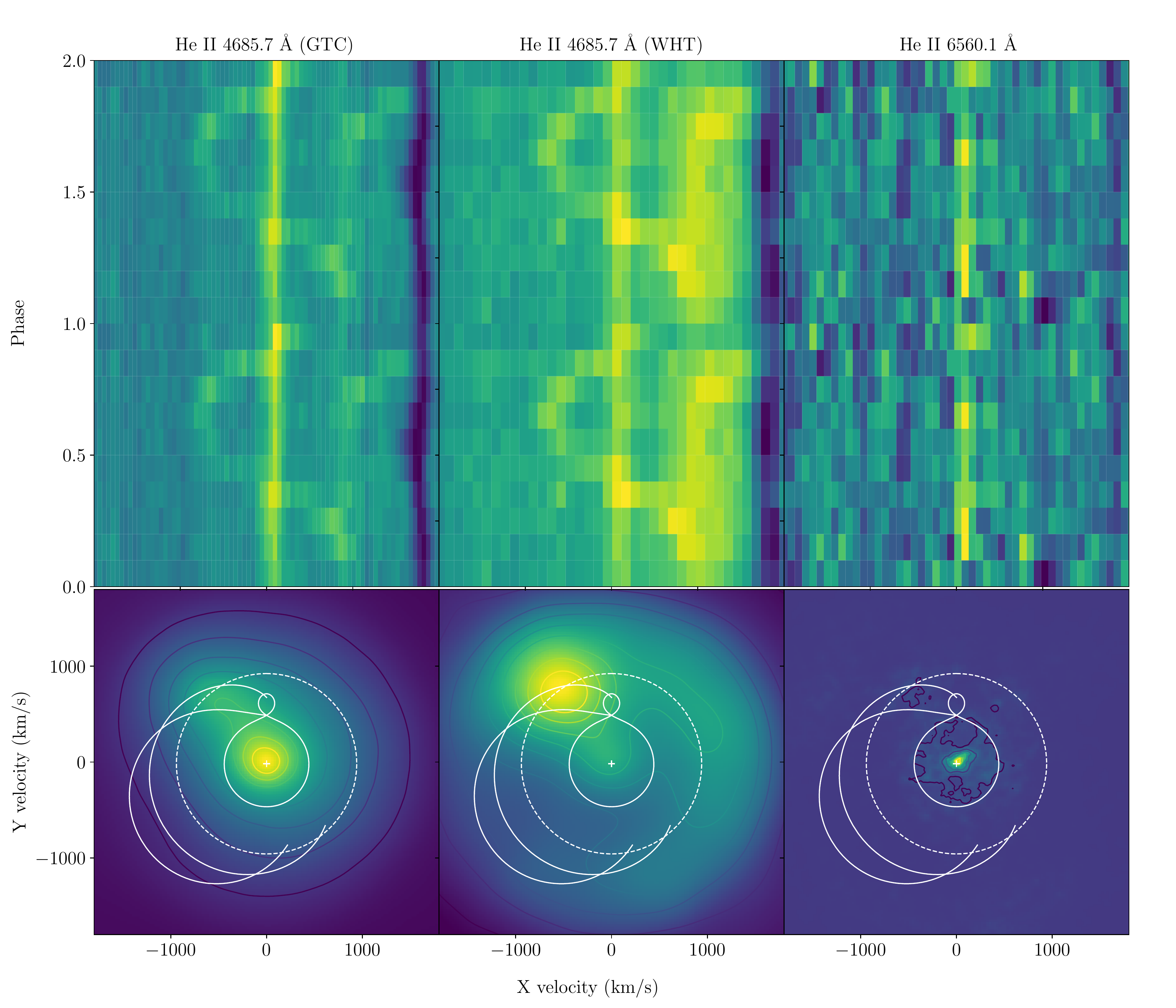}
    \caption{\textit{Above:} Phase-folded, trailed spectra of \heii\ lines using our data from the GTC (left panels) and WHT (centre and right panels). Absorption is shown in blue (dark for greyscale prints), and emission in yellow (light). The 4686\,\AA\ line shows a prominent, almost motionless central spike emission feature and an S-wave resulting from the bright spot (maximum blue excursion at phase 0.7). The 6560\,\AA\ line shows another central spike, albeit much weaker, while the bright spot is not visible.
    \textit{Below:} Doppler maps produced from the individual spectra. As in the trailed spectra, the 4686\,\AA\ Doppler maps show a central spike and one bright spot, and the 6560\,\AA\ Doppler map shows just the central spike.
    On the Doppler maps we also plot the predicted velocities of the Roche lobes (solid lines), the central white dwarf (cross), the stream of infalling matter (solid lines), and the velocity corresponding to the photometrically-measured bright spot separation radius (dotted line). The rotation of the map is such that the predicted position of the central white dwarf is below the origin of the map.
    }
    \label{fig:trail-gtc-g14}
\end{figure*}

\begin{table}
\caption{Measured positions in velocity space of the central spike and bright spot in the GTC \heii\ 4686\,\AA\ line of Gaia14aae, as measured using the method of shifting Gaussians. The systemic velocity offset of each component is denoted by $\gamma$.}
\begin{tabular}{lccc}
\hline
Feature & $K_x$ (km/s) & $K_y$ (km/s) & $\gamma$ (km/s) \\
\hline
Central Spike & 0.9 $\pm$ 3.8 & -13.8 $\pm$ 3.1 & 60.7 $\pm$ 2.4\\
Bright Spot & -548.9 $\pm$ 12.4 & 662.1 $\pm$ 5.8 & 16.0 $\pm$ 5.7\\
\hline
\end{tabular}
\label{tab:fitting-4686-g14}
\end{table}

\begin{table*}
\caption{Coordinates in velocity space of the bright spots in the \hei\ emission lines of Gaia14aae, measured from Doppler maps based on our GTC spectra using a two-dimensional Gaussian fit. \review{We also quote total velocity $K_\mathrm{tot}$ and phase $\phi$.} The systemic velocity in each map was assumed to be 16\,km/s.
}
\begin{tabular}{l | cc | cc | cc | cc}
\hline
Line &\multicolumn{4}{c}{Bright spot 1}&\multicolumn{4}{c}{Bright spot 2}\\
(\AA) & $K_x$ (km/s) & $K_y$ (km/s) & $K_\mathrm{tot}$ (km/s) & $\phi$ ($^\circ$) & $K_x$ (km/s) & $K_y$ (km/s) & $K_\mathrm{tot}$ (km/s) & $\phi$ ($^\circ$) \\
\hline
4471.5 & $-860 \pm 30$ & $550 \pm 40$ & $1020 \pm 40$ & $-123 \pm 2$ & $660 \pm 70$ & $-560 \pm 80$ & $860 \pm 70$ & $49 \pm 5$ \\
4713.1 & $-1280 \pm 80$ & $670 \pm 70$ & $1450 \pm 80$ & $-117 \pm 3$ & $0 \pm 200$ & $-1390 \pm 60$ & $1390 \pm 60$ & $0 \pm 8$\\
4921.9 & $-860 \pm 40$ & $680 \pm 30$ & $1100 \pm 40$ & $-129 \pm 2$ & $430 \pm 240$ & $-930 \pm 140$ & $1020 \pm 160$ & $25 \pm 13$\\
5015.7 & $-790 \pm 60$ & $580 \pm 50$ & $980 \pm 50$ & $-126 \pm 3$ & $170 \pm 80$ & $-950 \pm 20$ & $970 \pm 30$ & $10 \pm 5$\\
5047.7 & $-930 \pm 30$ & $640 \pm 40$ & $1130 \pm 30$ & $-124 \pm 2$ &  $-180 \pm 80$ & $-1170 \pm 30$ & $1180 \pm 30$ & $-9 \pm 4$\\
5875.6 & $-720 \pm 30$ & $420 \pm 80$ & $830 \pm 50$ & $-120 \pm 5$ & $340 \pm 170$ & $-690 \pm 100$ & $770 \pm 120$ & $26 \pm 12$ \\
\hline
\end{tabular}
\label{tab:fitting-dopp-g14}
\end{table*}

In Figures~\ref{fig:trail-wht-g14}-\ref{fig:trail-gtc-g14} we show trailed, phase-folded spectra using our data from the GTC and WHT. We show selected spectral lines in each case, including a combination of \hei, \heii, and \siii. 

For each spectral line we show a Doppler map \citep{Marsh1988}. 
Doppler maps are widely used in the analysis of CV and AM\,CVn systems \citep[eg.][]{Roelofs2005a,Roelofs2006,Kupfer2013,Kupfer2016}.
Such maps display the emission sources of each component of the spectral line in velocity space.
All motions are assumed to be in the plane of the binary system, with no vertical motion being considered.
These maps were produced from our spectra, after excluding all spectra that were affected by eclipses.
On the Doppler maps, we plot theoretical predictions for the Roche lobes, the central white dwarf, and the disc radius at which the bright spot is expected based on eclipse photometry. We also plot two versions of the stream of infalling matter: the ballistic velocities of the stream, and the Keplerian velocities corresponding to positions along the stream.
These Doppler maps are oriented such that the predicted velocity of the central white dwarf is in the negative y-direction.

The trailed spectra and Doppler maps of \hei\ and \siii\ lines are qualitatively similar to each other. Each shows a prominent bright spot which appears as an S-wave in the trailed spectra, and as an emission feature in the top left of the Doppler map. Both \hei\ and \siii\ lines also show a weaker second bright spot, offset by approximately 120$^\circ$ relative to the first bright spot and appearing at the bottom of the Doppler maps.
Finally, the absorption core of all \hei\ lines is clearly visible. In the trailed spectra, the absorption core appears to undergo radial velocity shifts. However, the phase of these velocity shifts is difficult to interpret as following the central white dwarf or the accretion disc. Instead, we interpret these apparent shifts as corresponding to interference from the two bright spots. 

The trailed spectra and Doppler maps of the \heii\ 4686\,\AA\ line are different to \hei. The first bright spot is still present, but a central spike feature is prominent while the second bright spot is not visible. Small velocity shifts can be seen in the central spike, though at a level much lower than the width of the feature itself (the velocity shifts are of order 10s of km/s, where the width of the feature is of order 100s km/s).
In the Doppler maps, the central spike feature is slightly offset in a downwards direction with respect to 0, consistent with the predicted velocity of the central white dwarf.


Two methods have been previously used to measure the velocity shifts of such features: a method of shifting Gaussians applied to the individual spectra \citep{Marsh1999,Kupfer2016}, and a method in which the Doppler map is fitted with a two-dimensional Gaussian \citep{Roelofs2005a,Roelofs2006,Kupfer2013,Kupfer2016}. For GP\,Com and V396\,Hya, using high resolution data, the two methods were found to give consistent results \citep{Kupfer2016}. 
For our data we found that optimum results could be obtained by using the method of shifting Gaussians to fit the \heii~4686\,\AA\ line, as the features are present with sufficient signal strength in individual spectra for the method to converge, and it returns results at a higher precision than the Doppler map fitting method.
In the \hei\ lines the bright spots (in particular the second bright spots) are weaker, and the method of shifting Gaussians was unable to converge on these features in individual spectra. We instead used the Doppler map fitting method for all \hei\ lines.

In the method of shifting Gaussians, the region of interest in each spectrum is simultaneously fitted, using a series of Gaussians whose velocity is offset from the rest wavelength by 
\begin{equation}
K_x \sin ( 2 \pi \phi ) + K_y \cos ( 2 \pi \phi ) + \gamma
\label{eq:shifting-gaussians}
\end{equation}
where $K_x$ and $K_y$ are the position of the feature in velocity space, $\phi$ is the orbital phase at which the spectrum was taken (defined such that eclipses occur at integer values of $\phi$), and $\gamma$ is any systemic Doppler shift of the feature from the rest wavelength. 

In order to estimate uncertainties a bootstrapping process was used. 2000 sets of bootstrapped spectra were produced from the parent set. Within each set, each spectrum was produced from its parent spectrum in the original set by selecting with replacement $N$ of the $N$ pixels in the parent spectrum. The method of shifting Gaussians was then applied to each of these bootstrapped sets of spectra. The results from this bootstrapping process were examined by eye, and found to be approximately normally distributed for each parameter. The final uncertainty was chosen as half the difference between the 16th and 84th percentile of these results.

For the Doppler map fitting method, we measured the positions of the two bright spots in velocity space by producing Doppler maps of the spectral lines and fitting the features with a two-dimensional Gaussian. 
Uncertainties on these values were derived using the bootstrapping method described by \citet{Wang2017,Wang2018}\cmnt{(Also Louise thesis if published in time?)}.
2000 sets of bootstrapped spectra were produced as described above.
After this, a Doppler map was produced from each set of bootstrapped spectra. 
Features of interest were fitted using a two-dimensional Gaussian in each bootstrapped Doppler map.
The uncertainty on each parameter was again chosen as half the difference between the 16th and 84th percentile of the bootstrapped results.

The results for the \heii\ 4686\,\AA\ line are shown in Table~\ref{tab:fitting-4686-g14}.
The total projected velocity for the central spike (CS) is $K_\mathrm{CS} = \sqrt{K_x^2 + K_y^2} = $~\kwd. This can be compared to our prediction of $K_\mathrm{WD} = 17.6 \pm 1.0$\,km/s based on the photometry-derived masses and inclination of \citet{Green2018}. The two values agree at the level of 1.1\,$\sigma$. 
We can also measure the orbital phase offset between the central spike and the predicted period of the central white dwarf by $\phi_\mathrm{CS} = \arctan(-K_x/K_y) =$~\phiwd, which agrees with the expected $0\,^\circ$ offset.

The systemic velocity of the bright spot in the \heii\ 4686\,\AA\ line, $\gamma = 16.0 \pm 5.7$\,km/s, agrees well with the velocity of $15.5 \pm 2.7$\,km/s measured from the wings of the \hei\ lines in Section~\ref{sec:rv}, and is likely due to the overall velocity of the binary. Relative to this velocity, the central spike is redshifted by a further $44.7 \pm 6.2$\,km/s. Similar redshifts have been seen in the central spikes of other AM\,CVns and can be attributed to a gravitational redshift. We discuss this further in Section~\ref{sec:central_spike}.

The positions of the \hei\ bright spots are shown in Table~\ref{tab:fitting-dopp-g14}. The measured positions of each bright spot is consistent between different emission lines, and the first spot is consistent with the bright spot in \heii. It can be seen from the Doppler maps that both spots appear to lie on or close to \review{either the velocity of the ballistic stream or the corresponding Keplerian velocity of its position}. We discuss this further in Section~\ref{sec:position_bs}.


\section{Discussion}
\label{sec:discussion}

\subsection{The Origin of the Central Spike}
\label{sec:central_spike}

The central spike appears to follow the central white dwarf in both $K$-velocity and orbital phase, and disappears when the central white dwarf is eclipsed. This confirms the widely used assumption that the spike originates on or near the surface of the white dwarf, and traces its motion \citep{Marsh1999,Kupfer2016}. 
This agreement can also be taken as an independent test of the stellar properties measured by \citet{Green2018}, producing a reasonably secure confirmation. 


As discussed in Section~\ref{sec:doppler}, the spike has a baseline redshift of $44.7 \pm 6.2$\,km/s relative to the velocity of the binary. Such redshifts have been previously seen in AM\,CVns.
The redshift of the central spike is a combination of two possible sources: gravitational redshift, and the Stark effect which causes blueshifts in \hei\ features. As the Stark effect does not affect \heii, our measured redshift is likely to be entirely gravitational in origin.

This gravitational redshift can be used to further constrain the origin site of the central spike. Given the mass and radius of the central white dwarf measured photometrically \citep[Table~\ref{tab:av-results}]{Green2018}, a redshift of $60 \pm 2$\,km/s would be expected at the white dwarf surface. The redshift we measure for the central spike disagrees with this. Instead our redshift implies that the origin site is a distance of $0.34^{+0.26}_{-0.17} R_\mathrm{WD}$ above the surface of the white dwarf. 
\review{
Line emission from this height is difficult to explain. 
The emitting material cannot be rotating with the Keplerian velocity as this would significantly broaden the emission line. 
Infalling material with a significant velocity would also produce Doppler broadening, and hydrostatic suspension at this height is difficult to believe for material cool enough to be dominated by singly ionised helium.
It is possible that the mass of the central white dwarf is significantly lower than the eclipse-based measurement; surface emission would suggest $M_1 = 0.75 \pm 0.05 M_\odot$ rather than $0.87 \pm 0.02 M_\odot$.
However, it is more likely that the redshift measurement is unreliable. The most likely source of error is the measured RV of the binary, which (as discussed in Section~\ref{sec:rv}) is subject to several possible complications. 
Finally, we note the large and non-Gaussian uncertainties on this measured height, which do not allow us to conclusively say that the emission site is inconsistent with the white dwarf surface.}


In many other AM\,CVns, the central spike is visible in both \hei\ and \heii. It may be that for these systems, either the overall disc is cooler or the emission site of the central spike is further from the white dwarf, resulting in a cooler emission site that allows both \hei\ and \heii\ to be seen. 
A cooler disc may be a result of a lower mass transfer rate or a lower mass white dwarf.
The mass transfer rate in Gaia14aae is known to be high for its orbital period \citep{Campbell2015,Ramsay2018}, but we note that it is still significantly lower than the mass transfer rate of AM\,CVn itself, in which \hei\ central spike emission is seen.
Alternatively, the absence of central spike emission from the \hei\ lines in Gaia14aae may be related to the edge-on disc absorption seen in this system.

\subsection{The Positions of the Bright Spots}
\label{sec:position_bs}

\begin{figure}
	\includegraphics[width=\columnwidth]{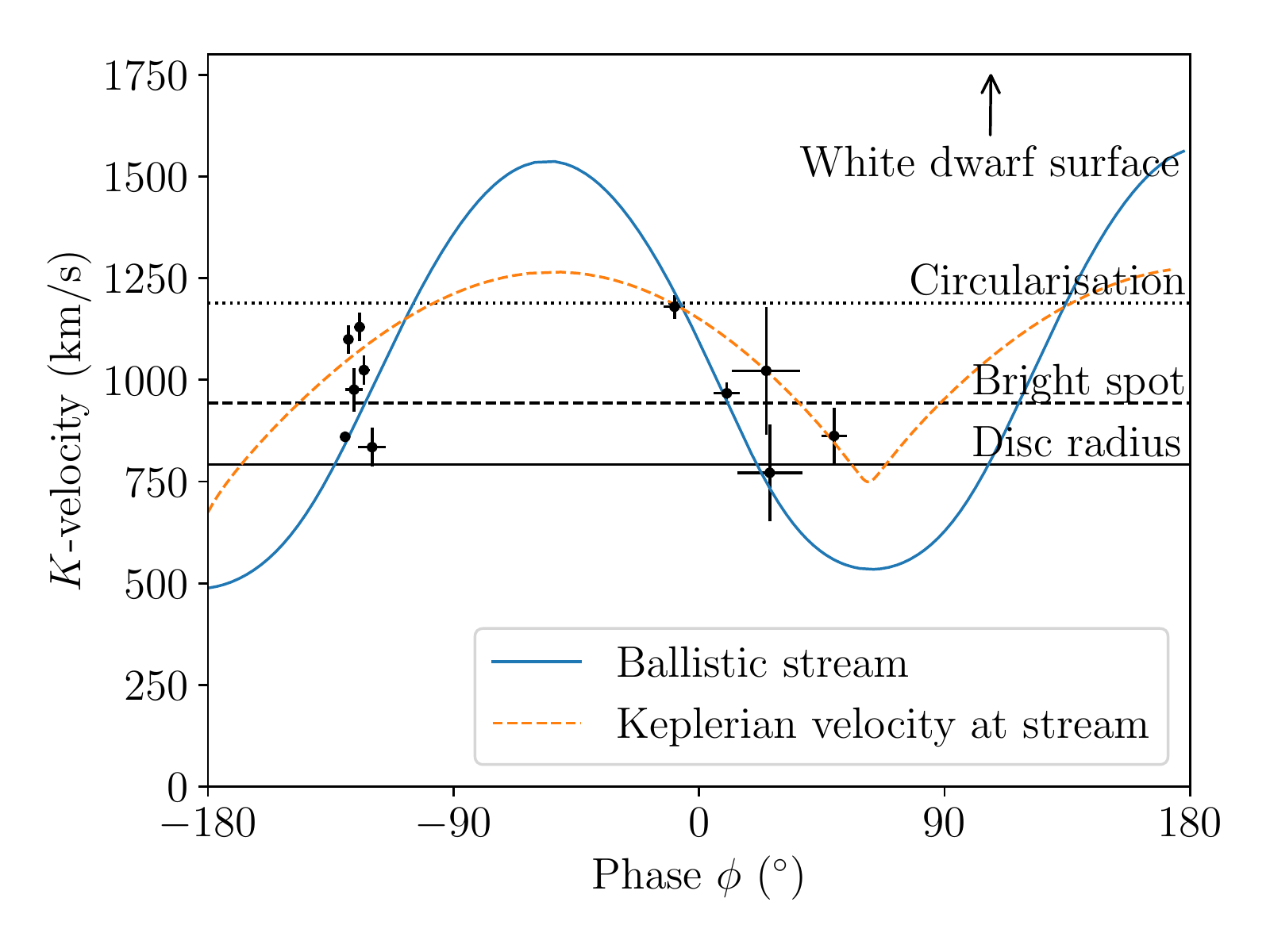}
    \caption{The measured velocities of the two bright spots, transformed from the values in Table~\ref{tab:fitting-dopp-g14} to absolute velocity $K$ and orbital phase $\phi$, are plotted as black points. We show for comparison the ballistic velocity of the infalling matter stream, and the Keplerian velocity at the position of each point in the stream. The latter may be seen as an approximation for the disc velocity at that point. Lastly, we show as horizontal lines the Keplerian velocities of three radii within the disc: the photometrically measured separation of the first bright spot and outer radius of the disc, and the circularisation radius. Note that larger radii will have a lower Keplerian velocity.
    All measurements for both bright spots seem to be consistent with the ballistic or Keplerian velocities of the stream, and are at a radius roughly consistent with the photometric measurement of the separation of the first bright spot.
    }
    \label{fig:kphi-g14}
\end{figure}

As can be seen in Figs.~\ref{fig:trail-wht-g14}-\ref{fig:trail-gtc-g14}, both bright spots in the \hei\ lines of Gaia14aae lie either on or close to the path of a ballistic stream of matter. In Fig.~\ref{fig:kphi-g14}, we plot the absolute velocity amplitude ($K$) and binary phase angle ($\phi$) of the bright spots in each He line in the GTC data. 
We do not use the \hei\ 4713\,\AA\ measurements, which are clearly discrepant with measurements from other lines, perhaps due to blending between \hei\ 4713\,\AA\ and \heii\ 4686\,\AA.
We define $\phi$ from the Doppler maps as the rotation of the component anticlockwise around the centre, with the accreting white dwarf at phase zero.
In Fig.~\ref{fig:kphi-g14} we additionally plot lines corresponding to the velocity and phase of a ballistic stream of matter \citep{Lubow1975}, and the Keplerian velocity that the disc would be expected to have at that position.

The first bright spot originates from the point at which the infalling stream of matter is decelerated to match the velocity of the accretion disc. As such its velocity is expected to be either equal to that of the ballistic stream, equal to the Keplerian disc velocity at the point of collision, or between the two. The scatter in our measurements appears to be centred between the two velocities, but our most precise measurement, the \heii\ bright spot, lies exactly on the velocity of the ballistic stream. 
This is of interest for AM\,CVn mass ratio derivations. In order to convert from the projected velocity of the white dwarf ($K_{WD}$) to the true velocity of the white dwarf, from which the mass ratio can be found, one must first estimate the inclination of the system. 
The inclination can be measured from the velocity of the bright spot, if it is assumed to agree with either the ballistic stream or the Keplerian disc velocity. Past studies have generally either assumed the ballistic case \citep{Roelofs2006} or used the ballistic and Keplerian cases to derive lower and upper limits \citep{Marsh1999,Kupfer2016}.
Based on the scatter in our measurements, the latter assumption is safer.

While a second bright spot is seen in many AM\,CVns \citep{Roelofs2005a,Roelofs2006,Kupfer2013,Kupfer2016}, its origin is still unclear.
Past studies have suggested it may result from some fraction of the infalling stream continuing past its first contact with the disc, and instead interacting with the disc at a later point in its path \citep[eg.][]{Kupfer2016}. \citet{Roelofs2005a} suggested an explanation in which this `overflowing' fraction of the stream leaves the disc along its ballistic trajectory and causes the second bright spot as it re-enters the disc. 
An alternate proposal has been modelled by several numerical simulations \citep[Wood, priv.\ comm.]{Armitage1998}. In this proposal, the impact of the stream at the edge of the accretion disc causes some fraction of the infalling material to deflect above and below the disc. The deflected material continues to follow an approximately ballistic path, and terminates at a second bright spot where the material reimpacts the disc.
Other suggested models include those in which a standing wave forms in the outer disc, potentially explaining the approximate 120$^\circ$ separation of the two spots.

In Gaia14aae, the agreement of the second spot velocity with the path of the ballistic stream is suggestive. However, unlike in the mechanism suggested by \citet{Roelofs2005a}, the second bright spot in Gaia14aae appears at a region in which the ballistic stream would be travelling \textit{outward} through the disc, not inward. 
As such, the second bright spot in Gaia14aae may favour the mechanism of upward and downward deflection suggested by \citet{Armitage1998}, in which the path of the material is approximately ballistic.
However, we note that the location of the second bright spot in Gaia14aae does not exactly match the predictions of \citet{Armitage1998}, in that they predict a location near the closest approach of the stream to the central white dwarf. 
Indeed, it is unclear that the ballistic stream path remains physically meaningful beyond the point of closest approach.
It is interesting to note that the radius corresponding to the velocity of the second spot in Gaia14aae is approximately equal to the radius of the first bright spot, but it is unclear whether there is a physical reason for this.

If the second bright spot can be assumed to lie on the ballistic stream path in all AM\,CVns, this would provide an additional constraint on spectroscopic mass ratio measurements. However, given the uncertainty that remains regarding the formation mechanism of the second bright spot, extreme caution should be exercised before adopting such assumptions. It is possible that the agreement of our measured velocity and phase with the ballistic stream is entirely coincidental.

\subsection{Implications for the Formation of Gaia14aae}

There are three proposed channels by which AM\,CVns can form, with the relative importance of each channel being uncertain. 
In the white dwarf donor channel, both stars in a close binary evolve into white dwarfs via two stages of common envelope evolution. Gravitational wave radiation then drives the white dwarfs into contact, forming an AM\,CVn with a white dwarf donor \citep{Paczynski1967,Deloye2007}.
The helium star donor channel is similar, but differs in that the donor is a stripped giant core that is still undergoing helium fusion at the point of contact \citep{Savonije1986,Iben1987,Yungelson2008}.
In the evolved CV channel, the binary first becomes a hydrogen-accreting CV with an evolved (helium core) donor. Mass transfer removes the outer layers of the donor, exposing its helium core, and the accreted material becomes helium-dominated \citep{Tutukov1985,Podsiadlowski2003,Goliasch2015}.

A key purpose of characterising AM\,CVn binaries is to identify their formation channel. For Gaia14aae the results to date are ambiguous. 
The donor mass and radius of Gaia14aae agree well with models for the evolved CV channel \citep{Green2018}. 
They also lie in a region of mass-radius space which the helium star donor channel may be able to explain following tweaks to the model.

The biggest discrepancy with the evolved CV channel is that these models predict an amount of hydrogen in the transferred material generally on the order of 1\% \citep[Nelson, private communication, 2016;][]{Goliasch2015}. 
Smaller amounts of hydrogen are possible, but the initial conditions required become increasingly improbable as the hydrogen fraction decreases.
As the region of possible initial conditions becomes smaller, we would expect to find an increasing number of CVs below the period minimum with spectroscopically visible hydrogen and helium for every AM\,CVn that is found.

The upper limit on H$\alpha$ EW measured in Section~\ref{sec:hydrogen} is less constraining than was assumed by \citet{Green2018}. 
Allowing the accreted material to contain a higher fraction of hydrogen makes the evolved cataclysmic variable channel more favourable for Gaia14aae.

The abundances of CNO elements can also discriminate between evolutionary channels. In the white dwarf donor and evolved CV channels, the relative abundances of these elements will roughly match equilibrium abundances for the CNO cycle \citep{Nelemans2010}. In the helium star donor channel, helium burning increases the abundance of triple-$\alpha$ products C and O relative to N \citep{Yungelson2008}. The extent of this abundance change depends on the length of time for which burning is able to occur before mass transfer starts. 
In Gaia14aae we have detected N and O absorption lines, but no C. This may be evidence against the helium star donor channel and in favour of the evolved CV channel. However, given the marginal strength of the N and O detections, the non-detection of C cannot be taken as truly significant. Further observations in ultraviolet or infrared may allow for a tighter constraint on the CNO composition.

In \citet{Ramsay2018}, a direct measurement of the mass transfer rate of Gaia14aae was made based on the absolute magnitude of the bright spot. Mass transfer rate can be used to explore the donor's nature, as it constrains the donor's response to mass loss $\xi = d \log (R_2 / R_\odot) / d \log (M_2 / M_\odot)$. The measured mass transfer rate $\log \left( \dot{M} / \left[ M_\odot \mathrm{yr}^{-1} \right] \right) = -10.74 \pm 0.07$ implies $\xi = -0.31^{+0.23}_{-0.20}$. This value agrees better with the prediction based on the helium star and white dwarf donor channels ($\approx$ 0.2) than the prediction based evolved CV channel ($\approx$ 0.0). 
However, the measurement is not precise enough to rule out any channel. 
A more precise measurement of mass transfer rate may be possible by eclipse timing. AM\,CVn binaries show measurable increases to their orbital period over a timescale of years \citep{Copperwheat2011a,deMiguel2018}, governed by a combination of gravitational wave emission and mass transfer \citep{Deloye2007}. If the stellar masses are known, the rate of orbital period change may be used to measure mass transfer rate. Timing the eclipses of Gaia14aae will make it possible to measure the mass transfer rate to enough precision to discriminate between, or at least further constrain, the evolutionary channels.

There is growing evidence that Gaia14aae may have an unusual formation channel. When compared to other AM\,CVns, the donor of Gaia14aae appears to be unusually inflated for its orbital period, and indeed has the largest radius of all AM\,CVn donors with measured radii \citep[see Fig.~11 of][]{Green2018b}. It remains the second-longest period AM\,CVn to undergo a dwarf nova outburst \citep{Campbell2015}, and its absolute magnitude is unusually bright for systems of its orbital period \citep[][Fig.~3]{Ramsay2018}, both suggesting an unusually high mass transfer rate for its orbital period. The mass transfer rate measured in \citet{Ramsay2018} agrees with this, though with large uncertainties.

If Gaia14aae is atypical for a long-period AM\,CVn, this strengthens the case for its having formed by the evolved CV channel, which is expected to contribute a minority of AM\,CVns at long orbital periods. Binaries formed by the evolved CV channel are expected to have more inflated donors and higher mass transfer rates than AM\,CVns formed by the other two channels. 
A number of CVs with visible hydrogen have been detected that are below the typical CV period minimum and show both hydrogen and helium. Ten such systems are known \citetext{as of \citealp{Breedt2015}; see \citealp{Augusteijn1993,Augusteijn1996,Breedt2012,Breedt2014,Carter2013,Littlefield2013} for individual systems}.
As such, we might expect a small number of AM\,CVns to have formed by the same channel.


\section{Conclusions}

We have presented phase-resolved spectroscopy of Gaia14aae, the first known fully-eclipsing AM\,CVn. 
The system shows a series of \hei\ and \heii\ emission lines, as well as absorption and emission from several metals. 
The \hei\ lines have a line profile consistent with emission from a disc and two bright spots, while the \heii\ lines show one bright spot and the `central spike' emission feature seen in many AM\,CVns. 

By comparing in-eclipse and out-of-eclipse spectra, we confirm that the central spike is eclipsed at the same time as the central white dwarf. 
Using the Doppler shifts of the central spike, we derive a projected white dwarf velocity amplitude which is in agreement with our prediction based on the stellar properties derived from eclipse photometry, thus confirming the widely used but never directly tested assumption that this emission traces the velocity of the white dwarf.

We tested two methods to measure the radial velocity of the central spike. The widely used method of fitting to Doppler tomography was found to be too imprecise to measure the white dwarf velocity for Gaia14aae, but our method of sinusoidally shifting Gaussians produced a higher precision measurement. 

We measured the systemic velocities of both the central spike and the bright spot, using the difference to compute the gravitational redshift of the central spike. 
We found a gravitational redshift of $44.7 \pm 6.2$\,km/s, corresponding to an emission site $0.34^{+0.26}_{-0.17} R_\mathrm{WD}$ above the white dwarf surface. The temperature of this emission site is predicted to be $15600^{+700}_{-1200}$\,K, and it coincides with the peak of the temperature profile of the accretion disc.

The positions of both bright spots coincide with predicted velocities of the stream of matter. In the case of the first bright spot, this matches with expectations. In the case of the second bright spot, this is somewhat surprising as the spot appears on the outward trajectory of the accretion stream. This may suggest that the second spot, like the first, results from an interaction between stream and disc. However, the mechanism of such an interaction, and why it appears at this particular point on the stream's path, is unknown, and the association may be down to chance.

Our previous work on Gaia14aae \citep{Green2018} attempted to answer the question of how the binary formed. In that work we were unable to distinguish between two possible formation channels, the helium star donor channel and the evolved cataclysmic variable channel. 
In this study we have presented marginal detections of nitrogen and oxygen compared to a non-detection of carbon, and an upper limit on hydrogen that was less restrictive than expected.
We also discuss evidence from \citet{Campbell2015} and \citet{Ramsay2018} that Gaia14aae may have an unusually high mass transfer rate for its orbital period.
Combining these points, we tentatively suggest that Gaia14aae may be an unusual example of an AM\,CVn that has formed by the evolved cataclysmic variable channel. However, these arguments are far from conclusive.
We recommend further observations that may help to constrain the formation channel of Gaia14aae, including ultraviolet or infrared \review{spectroscopy to search for carbon}, and long-term timing of its eclipses \review{in order to constrain the gravitational wave radiation and hence its mass transfer rate}.


\section*{Acknowledgements}

The authors would like to thank Matt Wood for helpful discussion regarding modelling of the second bright spot, and Chris Manser for discussion regarding the Doppler maps. We would also like to thank the anonymous reviewer for the comments and suggestions.
\cmnt{Any other discussions, reviewer?}

MJG acknowledges funding from an STFC studentship via grant ST/N504506/1. 
TRM and DTHS acknowledge STFC via grant ST/P000495/1.
EC has received funding from the European Research Council under the European Union's Horizon 2020 research and innovation programme (grant agreement no. 677706 -- WD3D).
\cmnt{Other funding numbers?}

The data reduction presented in this work was carried out primarily using \texttt{\sc Pamela} and \texttt{\sc Molly}, as well as various packages included in \texttt{\sc Starlink}. The analysis made use of \texttt{\sc python} packages including \texttt{\sc numpy}, \texttt{\sc matplotlib}, \texttt{\sc scipy}, and \texttt{\sc astropy}.

This work is partly based on observations made with the William Herschel Telescope (WHT). The WHT is operated on the island of La Palma by the Isaac Newton Group of Telescopes in the Spanish Observatorio del Roque de los Muchachos of the Instituto de Astrof\'{i}sica de Canarias.
This work is also partly based on observations made with the Gran Telescopio Canarias (GTC), installed in the Spanish Observatorio del Roque de los Muchachos of the Instituto de Astrof\'{i}sica de Canarias, in the island of La Palma.




\bibliographystyle{mnras}
\bibliography{refs} 








\bsp	
\label{lastpage}
\end{document}